\documentstyle[12pt]{article}

\begin{document}
\baselineskip 7.5mm
\begin{flushright}
\begin{tabular}{l}
CERN-TH/97-178 \\
hep-ph/9707428
\end{tabular}
\end{flushright}

\vspace{4mm}

\begin{center}
{\Large \bf Reply to} {\Large 
\it Comment on ``Pulsar velocities and neutrino
oscillations'' } \\
\vspace{3mm}
Alexander Kusenko$^*$ and Gino Segr\`e$^{**}$ \\
\vspace{2mm}
{\small \it $^*$Theory Division, CERN, 1211 Geneva, Switzerland} \\
{\small \it $^{**}$Department of Physics and Astronomy, University of 
Pennsylvania, Philadelphia, PA 19104-6396 , USA }

\end{center}

The explanation to pulsar birth velocities proposed in Ref.~\cite{ks} 
assumes that neutrino oscillations take place in the vicinity of the
neutrinospheres, where the matter  density is $\rho = 10^{10} -
10^{12}$ g/cm$^3$ \cite{s}.  A $\Delta k/k \sim 1\% $ asymmetry in the
momentum distribution of outgoing neutrinos can explain the observed motion
of pulsars.  We evaluated $\Delta k/k$ using the 
approximation $dN_e/dT \approx (\partial N_e/\partial T)_{\mu_e}$ 
and obtained 

\begin{equation}
\frac{\Delta k}{k} =  0.01 
\left ( \frac{B}{3 \times 10^{14} {\rm G}} \right ).
\label{dk0}
\end{equation} 
In this approximation the total kick does not depend on $Y_e$ or $\rho $.

An alternative approximation was proposed in the preceding Comment
\cite{comment}.  Replacing $d N_e/dT $ by its value at constant $Y_e$ 
yields an upper estimate of

\begin{equation}
\frac{\Delta k}{k} = 0.01 \left (\frac{0.1}{Y_e} \right )^{2/3}
\left ( \frac{10^{11} {\rm g/cm}^3}{\rho} \right )^{2/3} 
\left ( \frac{B}{2 \times 10^{15} {\rm G}} \right ).
\label{dk}
\end{equation}

Neutrino oscillations must take place below the electron neutrinosphere, 
but above the $\tau $ neutrinosphere.  (We note in passing that several 
different definitions of neutrinosphere are found in the literature and
refer the reader to Refs. \cite{s,c} for discussion.) 
For $\rho \sim (1-3)\times 10^{11} $ g/cm$^3$~\cite{c} and the
time average of $Y_e \approx 0.1$, 
Qian's approximation yields a somewhat higher
prediction for the magnetic field inside the neutron star.   

These two estimates are in reasonable agreement with each other, given the
uncertainties in the input parameters, the geometry of the magnetic
field and the simplified picture
of neutrino emission that comes with the notion of a neutrinosphere.

\end{document}